\documentclass[11pt]{article}
\usepackage{fa2025}
\usepackage{amsmath}
\usepackage{cite}
\usepackage{url}
\usepackage{graphicx}
\usepackage{color}
\usepackage{siunitx}
\usepackage[utf8]{inputenc}
\usepackage{subcaption}
\usepackage{placeins}

\title{Ambisonics Encoder for Wearable Array with Improved Binaural Reproduction}

\multauthor
{Yhonatan Gayer$^{1}$ \hspace{1cm} Vladimir Tourbabin$^2$ \hspace{1cm} Zamir Ben-Hur$^2$} { \bfseries{\hspace{1cm} David Alon$^2$ \hspace{1cm} Boaz Rafaely$^{1,2}$}\\
  $^1$ School of Electrical and Computer Engineering, Ben-Gurion University of the Negev, Beer-Sheva, Israel. \\
$^2$ Reality Labs Research at Meta, Redmond, WA, USA.
\correspondingauthor{gayery@post.bgu.ac.il}{Yhonatan Gayer et al.}
}

\begin{document}

\maketitle
\begin{abstract}

Ambisonics Signal Matching (ASM) is a recently proposed signal-independent approach to encoding Ambisonic signal from wearable microphone arrays, enabling efficient and
standardized spatial sound reproduction. However, reproduction accuracy is currently
limited due to the non-ideal layout of the microphones. This research introduces an enhanced
ASM encoder that reformulates the loss function by integrating a Binaural Signal
Matching (BSM) term into the optimization framework. The aim of this reformulation is to
improve the accuracy of binaural reproduction when integrating the Ambisonic signal with
Head-Related Transfer Functions (HRTFs), making the encoded Ambisonic signal better suited for binaural reproduction. This paper first presents the mathematical formulation developed to align the ASM and BSM objectives in a single loss function, followed by a simulation study with a simulated microphone array
mounted on a rigid sphere representing a head-mounted wearable array. The analysis
shows that improved binaural reproduction with the encoded Ambisonic signal can be
achieved using this joint ASM-BSM optimization, thereby enabling higher-quality
binaural playback for virtual and augmented reality applications based on Ambisonics.
\end{abstract}

\section{Introduction}
\label{sec:introduction}
Binaural reproduction is essential for creating immersive auditory experiences in virtual reality (VR) and augmented reality (AR), enabling spatial perception by enabling real-world acoustic cues \cite{Review-Paper}. Ambisonics \cite{book-Ambisonics} has emerged as a powerful method for spatial audio representation due to its ability to encode sound fields independent of the playback setup. When combined with individualized Head-Related Transfer Functions (HRTFs) \cite{HRTF_measurements}, Ambisonics allows for accurate binaural reproduction over headphones, preserving directionality and depth cues.

Conventionally, Ambisonic signal are captured using spherical microphone arrays \cite{PWD-with-Spherical-conv}. However, such capture rely on specialized arrays \cite{SH_Processing-book, em32}, limiting its use in mobile or wearable VR/AR applications \cite{Spherical-Ambisonics}.
Therefore, due to limitations in Ambisonics encoding, alternative binaural rendering approaches such as Binaural Signal Matching (BSM) have been proposed \cite{BSM-conference-paper}. These methods bypass the Ambisonics format and directly optimize the binaural signals to better match the target spatial cues. However, this approach is not as generic and versatile as Ambisonics, as it is both array and HRTF dependent.

To maintain the benefits of Ambisonics-based encoding while achieving performance that is close to BSM-based binaural reproduction, recent work has introduced the use of residual channels \cite{ASM}. This approach demonstrated that encoding additional channels beyond standard Ambisonics can achieve BSM-level accuracy.

Building on this concept, this paper aims to develop a First-Order Ambisonics (FOA) encoding method specifically optimized for binaural reproduction, but through modification of the encoding loss function, rather than through the addition of channels.

\section{Background}

This section provides the mathematical foundation for the signal model, presents the Ambisonics representation derived from this model, and outlines the formulations for encoding Ambisonics and rendering binaural signals using both spherical and arbitrary microphone arrays.

\subsection{Signal Model}

Consider an array consisting of $M$ omnidirectional microphones, where each microphone is located at spherical coordinates $(r_i, \theta_i, \phi_i)$ for $1 \leq i \leq M$. Where \(\theta_i\) represents the elevation angle and \(\phi_i\) represents the azimuth angle, and $r_i$ is a radius in meters $\forall 1 \le i \le Q$.
Additionally, assume the presence of $Q$ plane waves, each arriving from a distinct direction $(\theta_q, \phi_q)$, where $1 \leq q \leq Q$, forming the set $\Omega_Q$.
The array steering matrix, represented as $\mathbf{V} $, has dimensions $M \times Q$, where each element $[\mathbf{V} ]_{i,q}$ corresponds to the frequency response of the $i$-th microphone to a plane wave incident from $(\theta_q,\phi_q)$ at wave number $k$. The measured microphone signals are modeled as:
\begin{equation}
    \mathbf{x}(k)  = \mathbf{V}(k) \mathbf{s}(k)  + \mathbf{n}(k) 
    \label{eq:mic_model}
\end{equation}
where $\mathbf{x}(k)  = [x_1(k) , ..., x_M(k) ]^T$ is the vector of microphone signals, and each $x_i $ represents the measurement from the $i$-th microphone. The vector $\mathbf{s}(k)  = [s_1(k) , ..., s_Q(k) ]^T$ contains the source signals, with each entry representing the amplitude of a plane wave at the origin. The noise vector $\mathbf{n}(k)  = [n_1(k) , ..., n_M(k) ]^T$ consists of independent, identically distributed (i.i.d.) noise components, uncorrelated with $\mathbf{s}(k)$.

The binaural signal at the left or right ear can be computed using the spatial domain representation of the Head-Related Transfer Function (HRTF), denoted as $\mathbf{h}^{l,r}(k)  = [h^{l,r}(\theta_1,\phi_1,k), ..., h^{l,r}(\theta_Q,\phi_Q,k)]^T$, and the source vector $\mathbf{s}$ \cite{BSM_journal_paper}:
\begin{equation}
    p^{l,r}(k)  = \mathbf{h}^{l,r}(k) ^T \mathbf{s} (k)
    \label{eq:p_hs}
\end{equation}

\subsection{Ambisonics Representation}

The Ambisonics representation of the signal $\mathbf{s} $ with respect to the $Q$ plane waves can be written as \cite{SH_Processing-book}:
\begin{equation}
    \mathbf{a_{nm}}(k)  = \mathbf{Y}_{\mathbf{\Omega}_Q}^H \mathbf{s} (k)
    \label{eq:anm_Y_Hs}
\end{equation}
where $\mathbf{Y}_{\mathbf{\Omega}_Q}$ is the Spherical Harmonics (SH) matrix of size $Q \times (N_a+1)^2$. Each column vector $\mathbf{y_{nm}} = [Y_{nm}(\theta_1,\phi_1), ..., Y_{nm}(\theta_Q,\phi_Q)]^T$ represents SH functions of order $n$ and degree $m$ evaluated at $(\theta_q,\phi_q)$. The vector $\mathbf{a_{nm}}(k)  = [a_{00}(k) , ..., a_{N_aN_a}(k) ]^T$ contains the Ambisonics coefficients up to order $N_a$ and is of length $(N_a+1)^2$.

Binaural reproduction based on Ambisonics encoding is given by \cite{book-Ambisonics}:
\begin{equation}
    p^{l,r}(k)  = \tilde{\mathbf{h}}_{\mathbf{nm}}^{l,r} (k)^T \mathbf{a_{nm}} (k)
    \label{eq:p_hnm_anm}
\end{equation}
where $\tilde{\mathbf{h}}_{\mathbf{nm}}^{l,r} $ is a modified SH-domain representation of the HRTF. This representation is derived from the HRTF vector $\mathbf{h}_{\mathbf{nm}}^{l,r} $ using the relation $\tilde{h}_{nm}  = (-1)^m h_{n,-m} $. To ensure compatibility in Ambisonics reproduction, the orders of $\tilde{\mathbf{h}}_{\mathbf{nm}}^{l,r} $ and $\mathbf{a_{nm}} $ must be truncated such that $N_a = N_h$ \cite{Vlad-paper}.

For spherical arrays with near-uniform distributions, Ambisonics encoding can be performed accurately using Plane Wave Decomposition (PWD) \cite{PWD-paper}, ensuring correct computation of Ambisonics coefficients for all $n \leq N_a$ and $-n \leq m \leq n$, subject to the condition:
\begin{equation}
    (N_a + 1)^2 \leq M
    \label{eq:Na_sq_leq_M}
\end{equation}
This condition guarantees accurate Ambisonics encoding within the operational frequency range of the array \cite{rafaely2004analysis}.

\subsection{Ambisonics Encoding for Arbitrary Arrays}
\label{section:Ambisonics Encoding for Arbitrary Arrays}
To maintain conciseness, the wave number $k$ will be omitted in all sections of this paper from this point onward.
To enable binaural reproduction from arbitrary microphone arrays, Ambisonics channels can be directly encoded from the array signals. Traditionally, Ambisonics encoding is performed using spherical microphone arrays or purposefully designed configurations \cite{book-Ambisonics, SH_Processing-book, em32}. This formulation be can be generalized to arbitrary array geometries, as proposed in \cite{Vlad-paper, Parametric-ASM-like-paper}, which employs Tikhonov regularization \cite{Tikhinov}, thus providing a linear mapping from the microphone signals to the Ambisonics representation:
\begin{equation}
\begin{split}
    \hat{a}_{nm}&=\mathbf{c}_{nm}^H\mathbf{x},\\
    \forall \, 0 \leq n \leq &N_a, \, -n \leq m \leq n
\label{eq:anm_hat=Casmx}
\end{split}
\end{equation}
where $\hat{\mathbf{a}}_{\mathbf{nm}}  = [\hat{a}_{00} ,\dots,\hat{a}_{N_aN_a} ]^T$ denotes the estimated Ambisonics vector, of length $(N_a+1)^2$, and $\mathbf{c}_{nm}$ is the encoding Ambisonics filter of the same length.
This approach entails minimizing the following Normalized Mean Squared Error (NMSE) function to compute the optimal coefficients:
\begin{equation}
    \varepsilon_{nm}^{\text{ASM}} = E\left[\left\lVert \hat{a}_{nm}  - a_{nm} \right\rVert_2 ^2\right] \Big/ E\left[\left\lVert a_{nm}  \right\rVert_2 ^2\right]
    \label{eq:error_nm}
\end{equation}
Minimization is achieved by substituting \eqref{eq:mic_model}, \eqref{eq:anm_Y_Hs}, and \eqref{eq:anm_hat=Casmx} into \eqref{eq:error_nm}. For a signal-independent formulation, we assume that the noise $\mathbf{n} $ is white with covariance $\mathbf{R}_n = \sigma_n^2 \mathbf{I}$ and uncorrelated with $\mathbf{s} $, and that the source covariance matrix satisfies $\mathbf{R_s}  = \sigma_s^2\mathbf{I}$, corresponding to a diffuse sound field of $Q$ plane waves, the NMSE reduces to:
\begin{equation}
    \varepsilon_{nm}^{\text{ASM}} = \frac{\sigma_s^2 \left\lVert\mathbf{V} ^H\mathbf{c}_{nm}  - \mathbf{y}_{nm}\right\rVert_2^2 + \sigma_n^2 \left\lVert\mathbf{c}_{nm} \right\rVert_2^2}{\sigma_s^2\left\lVert\mathbf{y}_{nm}\right\rVert_2^2}
\label{eq:e_nm_AMB}
\end{equation}
Solving \eqref{eq:e_nm_AMB} leads to the optimal filter coefficients:
\begin{equation}
    \mathbf{c}_{nm}^{\text{ASM}} = [\mathbf{y}_{nm}]^H\mathbf{V} ^H
    \left(\mathbf{V}  \mathbf{V} ^H + \frac{\sigma_n^2}{\sigma_s^2}\mathbf{I}\right)^{-1}
\label{eq:cnm=ynmV^-1}
\end{equation}
where \( \mathbf{c}_{nm}^{\text{ASM}} \) represents the optimal filters that minimize (\ref{eq:e_nm_AMB}) with respect to \( [\mathbf{c}_{nm}]^H \).
Equation \eqref{eq:cnm=ynmV^-1} is valid provided that the inverse exists, which is ensured by the presence of the regularization term $\frac{\sigma_n^2}{\sigma_s^2}\mathbf{I}$. By substituting \eqref{eq:cnm=ynmV^-1} into \eqref{eq:anm_hat=Casmx}, an estimated Ambisonics representation is obtained via Ambisonics Signal Matching (ASM).

While this formulation is specific to each Ambisonics signal individually, it is extended here to reformulate the filter to encode all Ambisonics signals up to order \((N_a+1)^2\):
\begin{equation}
\begin{split}
    [\mathbf{C}_{\mathbf{nm}}^{\text{ASM}}]^H &= [\mathbf{c}_{00}^{\text{ASM}}, \dots, \mathbf{c}_{N_aN_a}^{\text{ASM}}]^H \\  
    &= [\mathbf{Y}_{\Omega_Q}]^H\mathbf{V} ^H
    \left(\mathbf{V}  \mathbf{V} ^H + \frac{\sigma_n^2}{\sigma_s^2}\mathbf{I}\right)^{-1}
    \label{eq:Cnm=ynmV^-1}
\end{split}
\end{equation}
This reformulation provides a more convenient representation of the ASM method, to be used later in this paper.

\subsection{Binaural Signal Matching (BSM) for Arbitrary Arrays}
\label{Sec:Binaural Signal Matching (BSM) for Arbitrary Arrays}
A direct approach for reproducing binaural signals from arbitrary microphone arrays is introduced in \cite{BSM_journal_paper}. This method directly derives the binaural signals from the microphone recordings via BSM. The estimated binaural signal is given by:
\begin{equation}
    \hat{p}^{l,r}_{\text{BSM}}  = [\mathbf{c}^{l,r}_{\text{BSM}}] ^H \mathbf{x} 
    \label{eq:p_bsm_cx}
\end{equation}
where $\mathbf{c}^{l,r}_{\text{BSM}} $ is the BSM filter, optimized by minimizing the Mean Squared Error (MSE):
\begin{equation}
    \varepsilon_{\text{BSM}} = E\left[\| \hat{p}^{l,r}_{\text{BSM}}  - p^{l,r}  \|_2^2 \right]
    \label{eq:e_bsm_mse}
\end{equation}
Under the assumptions of white noise $\mathbf{R}_n = \sigma_n^2 \mathbf{I}$ and an isotropic sound field where $\mathbf{R_s}  = \sigma_s^2\mathbf{I}$, the optimal BSM filter is obtained via Tikhonov regularization \cite{Tikhinov}:
\begin{equation}
    [\mathbf{c}^{l,r}_{\text{BSM}} ]^H = [\mathbf{h}^{l,r}] ^T \mathbf{V} ^H \left( \mathbf{V}  \mathbf{V} ^H + \frac{\sigma_n^2}{\sigma_s^2}\mathbf{I} \right)^{-1}
    \label{eq:c_bsm_h_v_inv}
\end{equation}
This formulation ensures robust binaural signal estimation for arbitrary array configurations.

\section{Proposed Method}
This paper introduces a unified approach for Ambisonics encoding by jointly minimizing ASM and BSM errors using a single loss function. Unlike existing methods that optimize these errors separately, the proposed method balances both objectives within a unified framework to compute an Ambisonic filter that encodes Ambisonics optimized for binaural reproduction.

\subsection{ASM Design for Binural Reproduction}
Binaural reproduction using encoded Ambisonics can be achieved by substituting (\ref{eq:Cnm=ynmV^-1}) into (\ref{eq:p_hnm_anm}):  
\begin{equation}
    \hat{p}^{l,r}_{\text{ASM}}  = [\tilde{\mathbf{h}}_{\mathbf{nm}} ]^T[\mathbf{C}_{\mathbf{nm}}^{\text{ASM}}] ^H \mathbf{x} 
    \label{eq:pASM=hnmCnmx}
\end{equation}
where $\hat{p}^{l,r}_{ASM} $ represents the binaural signal reproduced from encoded Ambisonics via ASM.
It is important to note that this formulation of the reproduced binaural signals does not minimize the binaural error. Instead, it represents the binaural signals reproduced using the encoded Ambisonics.

To derive the binaural error for the signal in (\ref{eq:pASM=hnmCnmx}), we can substitute $\hat{p}^{l,r}_{ASM} $ to replace $\hat{p}^{l,r}_{BSM} $ in (\ref{eq:e_bsm_mse}). By making the same assumptions of a diffuse sound field and white, uncorrelated microphone noise as in Sec. \ref{section:Ambisonics Encoding for Arbitrary Arrays}, we obtain:
\begin{equation}
\begin{split}
    \varepsilon_{\mathbf{nm}}^{\text{BSM}} &= \frac{\sigma_s^2 \left\lVert [\tilde{\mathbf{h}}_{\mathbf{nm}}^{l,r}]^T  [\mathbf{C}_{\mathbf{nm}}^{\text{ASM}}]^H\mathbf{V} - [\mathbf{h}^{l,r}]^T\right\rVert_2^2}{\sigma_s^2\left\lVert\mathbf{h}^{l,r}\right\rVert_2^2} \\
    &+ \hspace{9mm} \frac{\sigma_n^2\left\lVert[\tilde{\mathbf{h}}_{\mathbf{nm}}^{l,r}]^T  [\mathbf{C}_{\mathbf{nm}}^{\text{ASM}}]^H\right\rVert_2^2}{\sigma_s^2\left\lVert\mathbf{h}^{l,r}\right\rVert_2^2} 
\label{eq:e_bin_AMB}
\end{split}
\end{equation}
Although $\mathbf{C}_{\mathbf{nm}}^{\text{ASM}} $, as defined above, minimizes the error in (\ref{eq:e_nm_AMB}), it does not necessarily provide the coefficients that minimize (\ref{eq:e_bin_AMB}).
As a first step to formulating an Ambisonics encoder that minimizes the binaural error, the filter matrix $\mathbf{C}_{\mathbf{nm}}^{\text{ASM}}$ is restructured to the from of a long vector. Then, the product $[\tilde{\mathbf{h}}_{\mathbf{nm}}^{l,r}]^T  [\mathbf{C}_{\mathbf{nm}}^{\text{ASM}}]^H$ is replaced by the following equivalent term:
\begin{equation}
    [\tilde{\mathbf{h}}_{\mathbf{nm}}^{l,r}]^T  [\mathbf{C}_{\mathbf{nm}}^{\text{ASM}}]^H = [\mathbf{c}_{\mathbf{nm}, \text{flat}}^{\text{ASM}}]^H \mathbf{H}_{\mathbf{nm}}^{l,r}
    \label{eq:hnmCnm=cnmHnm}
\end{equation}
where 
\begin{equation}
    \mathbf{c}_{\mathbf{nm}, \text{flat}}^{\text{ASM}} = \bigg[\big[\mathbf{c}_{00}^{\text{ASM}}\big]^T, \dots, \big[\mathbf{c}_{N_aN_a}^\text{ASM}\big]^T\bigg]^T
    \label{eq:cnmASM_flat=[c00,..,cNaNa]^T}
\end{equation}
 is a vectorized form of the matrix \(\mathbf{C}^{\text{ASM}}_{\mathbf{nm}}\), changing its dimensions from \(M \times (N_a+1)^2\) to \(M(N_a+1)^2 \times 1\), and
 \begin{equation}
     \mathbf{H}_{\mathbf{nm}}^{l,r} = \big[ \tilde{h}_{00}^{l,r} \hspace{1mm}\mathbf{I},\dots, \tilde{h}_{N_aN_a}^{l,r}\hspace{1mm}\mathbf{I} \big]^T
     \label{eq:Hnm=[h00I,...,hNaNaI]^T}
 \end{equation}
is a Matrix of size $M(N_a+1)^2\times M$, where $\mathbf{I}$ is the unit matrix of size $M \times M$. 
Substituting (\ref{eq:cnmASM_flat=[c00,..,cNaNa]^T}) and (\ref{eq:Hnm=[h00I,...,hNaNaI]^T}) in (\ref{eq:hnmCnm=cnmHnm}) leads to equalities in these equations.
Now, (\ref{eq:e_bin_AMB})  can be rewritten using (\ref{eq:hnmCnm=cnmHnm}) and by further replacing $\mathbf{c}_{\mathbf{nm}, \text{flat}}^{\text{ASM}}$ with $\mathbf{c}_{\mathbf{nm}, \text{flat}}$, to denote that it is a free parameter to be optimized:
\begin{equation}
\begin{split}
    \varepsilon_{\mathbf{nm}}^{\text{BSM}} &= \frac{\sigma_s^2 \left\lVert [\mathbf{c}_{\mathbf{nm}, \text{flat}}]^H\mathbf{H}_{\mathbf{nm}}^{l,r}\mathbf{V} - [\mathbf{h}^{l,r}]^T\right\rVert_2^2}{\sigma_s^2\left\Vert\mathbf{h}^{l,r}]^T\right\rVert_2^2}\\
    &+ \hspace{2mm} \frac{\sigma_n^2 \left\lVert [\mathbf{c}_{\mathbf{nm}, \text{flat}}]^H \mathbf{H}_{\mathbf{nm}}^{l,r} \right\rVert_2^2}{\sigma_s^2\left\Vert\mathbf{h}^{l,r}]^T\right\rVert_2^2}
\label{eq:e^_bin_AMB}
\end{split}
\end{equation}
%
%
%
%
%
%
Now, $\mathbf{c}_{\mathbf{nm}, \text{flat}}^{\text{BSM}}$, can be computed as a solution to the minimization of (\ref{eq:e^_bin_AMB}), which can be achieved using Tikhonov regularization \cite{Tikhinov}:
\begin{equation}
\begin{split}
    [\mathbf{c}_{\mathbf{nm}, \text{flat}}^{\text{BSM}}]^H =& 
      \arg\min_{\mathbf{c}_{\mathbf{nm}, \text{flat}}} \varepsilon_{\mathbf{nm}}^{\text{BSM}} \\
    = & [\mathbf{h}^{l,r}]^T [\mathbf{H}_{\mathbf{nm}}^{l,r}\mathbf{V}]^H \times \\
    &\bigg(\mathbf{H}_{\mathbf{nm}}^{l,r}\mathbf{V} [\mathbf{H}_{\mathbf{nm}}^{l,r}\mathbf{V}]^H + \frac{\sigma_n^2}{\sigma_s^2}\mathbf{H}_{\mathbf{nm}}^{l,r}[\mathbf{H}_{\mathbf{nm}}^{l,r}]^H\bigg)^{-1}
\end{split}
\end{equation}
where, similar to (\ref{eq:cnmASM_flat=[c00,..,cNaNa]^T}), and (\ref{eq:Cnm=ynmV^-1}), $ [\mathbf{c}_{\mathbf{nm}, \text{flat}}^{\text{BSM}}]$ can be rearenged to the same form as in (\ref{eq:e_bin_AMB}):
\begin{equation}
    \mathbf{C}_{\mathbf{nm}}^{\text{BSM}} = \big[ \mathbf{c}_{00}^{\text{BSM}}, \dots, \mathbf{c}_{N_aN_a}^{\text{BSM}} \big]
\end{equation}

\subsection{Joint ASM-BSM design}
The authors of this paper demonstrated in \cite{ASM} that when the encoded Ambisonics order is small, i.e. $N_a \ll N_h$, the minimization of (\ref{eq:e_nm_AMB}) typically leads to high binaural error.
To address this limitation, this paper proposes minimizing the joint ASM-BSM error, defined as:  
\begin{equation}
    \varepsilon_{\mathbf{nm}}^{\text{joint}} = \alpha \sum_{n=0}^{N_a} \sum_{m=-n}^{n} \varepsilon_{nm}^{\text{ASM}} + (1-\alpha) \varepsilon_{\mathbf{nm}}^{\text{BSM}}
    \label{eq:e_joint}
\end{equation}  
where \( \varepsilon^{\text{joint}}_{\mathbf{nm}} \) represents the objective error function, and \( \alpha \) acts as a balancing factor. Both \( \varepsilon_{nm}^{\text{ASM}} \) as in (\ref{eq:e_nm_AMB}) and \( \varepsilon_{\mathbf{nm}}^{\text{BSM}} \) as in (\ref{eq:e^_bin_AMB}) depend on the Ambisoniocs encoder coefficients, as the variables to be optimized. The solution to the error in (\ref{eq:e_joint}) is a linear combination of the solutions to (\ref{eq:e_bin_AMB}) and (\ref{eq:e_nm_AMB}), expressed as:  
\begin{equation}
    \mathbf{C}_{\mathbf{nm}}^{\text{joint}} = \alpha \hspace{1mm} \mathbf{C}_{\mathbf{nm}}^{\text{ASM}} + (1-\alpha)\hspace{1mm}\mathbf{C}_{\mathbf{nm}}^{\text{BSM}}
    \label{eq:Cjoint}
\end{equation}
where $\mathbf{C}_{\mathbf{nm}}^{\text{joint}}$ represents the joint ASM-BSM filter.


\section{Experiment}
In this experiment, the joint ASM-BSM model is evaluated by measuring the ASM and BSM errors. The model is tested with different $\alpha$ values, allowing us to analyze the trade-off between the ASM and BSM solutions, highlighting the benefits of their combination.

\subsection{Setup}
Steering functions and HRTF were simulated in Matlab based on a spherical head model. A rigid sphere of radius $10\,$cm was simulated, and the ear and microphones were positioned on its surface as detailed below. Ambisonics of order $N_a=1$ was chosen for encoding, with a reference order of $N_a=20$. Far-field sources were simulated at 240 directions distributed nearly-uniformly in the entire directional space.

\begin{figure}[b]
    \centering
    \includegraphics[width=\columnwidth]{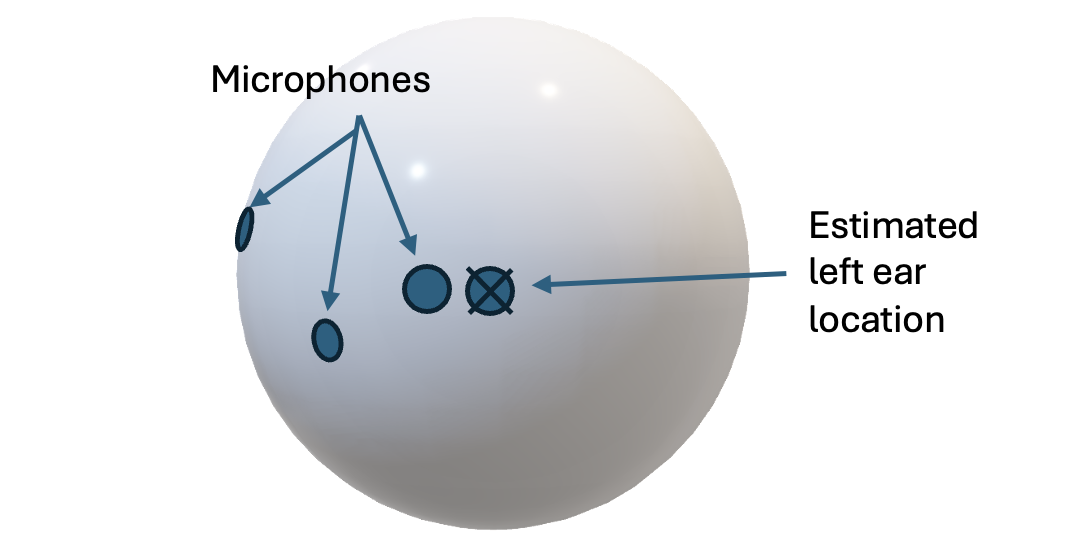}
    \caption{\fontsize{9}{11}\selectfont An illustration of the array from a horizontal view. The microphone positions on a rigid sphere, given in spherical coordinates $(\theta, \phi)$, are \( \left\{ (90^{\circ}, -70^{\circ}), (72^{\circ}, -35^{\circ}), (108^{\circ}, 0^{\circ}), (72^{\circ}, 35^{\circ}), (90^{\circ}, 70^{\circ}) \right\} \), and the array is symmetric. The ears are located at \( (90^{\circ}, \pm 90^{\circ}) \).}

    \label{fig:array plot}
\end{figure}

The microphone array consists of \( M = 5 \) microphones arranged along a semi-circle on the rigid sphere. The location of each microphone is defined in spherical coordinates \((\theta, \phi)\). The microphone positions are given by:  
$\{ (90^{\circ}, -70^{\circ}), (72^{\circ}, -35^{\circ}), (108^{\circ}, 0^{\circ}), (72^{\circ}, 35^{\circ})$, $ (90^{\circ}, 70^{\circ}) \}$ 
as illustrated in Fig. \ref{fig:array plot}. These positions aim to represent an array mounted on glasses, such as the one used for the EasyCom database \cite{donley2021easycom}.

\subsection{Methodology}
Joint ASM-BSM filters were as follows:
\begin{itemize}
    \item Joint - ASM: Joint design with $\alpha=1$, optimizing only for Ambisonics.
    \item Joint - BSM: Joint design with $\alpha=0$, optimizing only for BSM.
    \item Joint - ASM-BSM: Joint design with $\alpha=0.5$, balancing between ASM and BSM.
    \item Std BSM: The standard BSM approach as described in Sec. \ref{Sec:Binaural Signal Matching (BSM) for Arbitrary Arrays}.
\end{itemize}    
\subsection{Performance Measures}

To avoid confusion between the errors used for filter design and those used for performance evaluation, we define the relevant error measures separately in this subsection.
The per-channel error for the ASM, similar to (\ref{eq:e_nm_AMB}) is defined as:
\begin{equation}
    \xi_{nm}^{\text{ASM}} = \frac{\left\lVert\mathbf{V} ^H\mathbf{c}_{nm}^{\text{joint}}  - \mathbf{y}_{nm}\right\rVert_2^2}{\left\lVert\mathbf{y}_{nm}\right\rVert_2^2}
    \label{eq:xi_nm_ASM}
\end{equation}
This represents the reconstruction error for the $n,m$ Ambisonics channel using filter \( \mathbf{c}_{nm}^{\text{joint}} \) which is one of the columns of filter $\mathbf{c}_{nm}^{\text{joint}}$ corresponding to the $n,m$ channel.
The error for the binaural reproduction when jointly designing filters, similar to (\ref{eq:e^_bin_AMB}) is given by:
\begin{equation}
    \xi_{\mathbf{nm}}^{\text{BSM}} = \frac{\left\lVert [\tilde{\mathbf{h}}_{\mathbf{nm}}^{l,r}]^T  [\mathbf{C}_{\mathbf{nm}}^{\text{joint}}]^H\mathbf{V} - [\mathbf{h}^{l,r}]^T\right\rVert_2^2}{\left\lVert\mathbf{h}^{l,r}\right\rVert_2^2}
    \label{eq:xi_nm_BSM}
\end{equation}
where \( \mathbf{C}_{\mathbf{nm}}^{\text{joint}} \) is computed as in Eq. (\ref{eq:Cjoint}).
Serving as a reference for binaural reproduction, the performance of the Standard BSM is evaluated using:
\begin{equation}
    \xi^{\text{BSM}} = \frac{\left\lVert [\mathbf{c}^{l,r}_{\text{BSM}}]^H\mathbf{V} - [\mathbf{h}^{l,r}]^T\right\rVert_2^2}{\left\lVert\mathbf{h}^{l,r}\right\rVert_2^2}
    \label{eq:xi_BSM_MSE}
\end{equation}
where \( \mathbf{c}^{l,r}_{\text{BSM}} \) represents the standard BSM filter as in (\ref{eq:c_bsm_h_v_inv}).

\subsection{Results} 
\begin{figure}[t]
    \centering
    \includegraphics[width=0.45\textwidth]{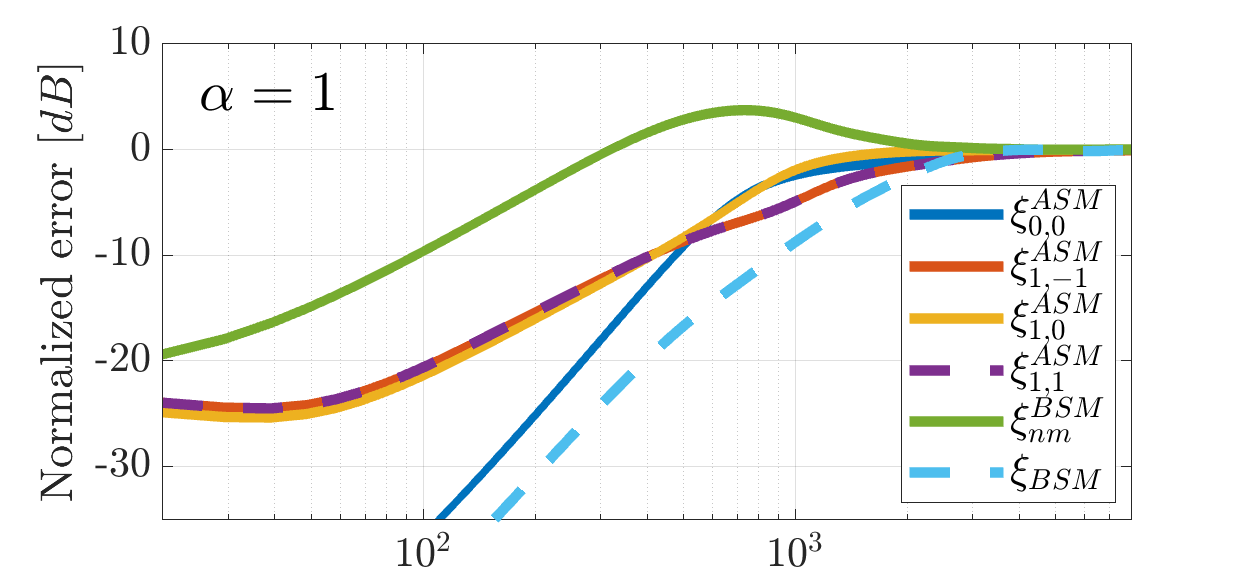}
    \includegraphics[width=0.45\textwidth]{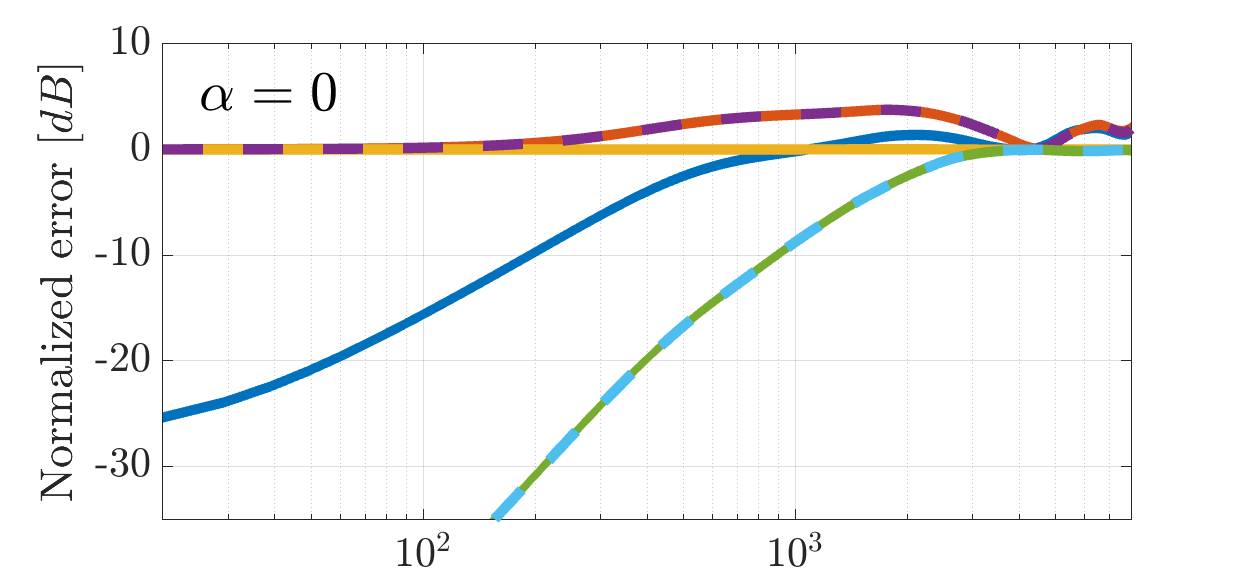}
    \includegraphics[width=0.45\textwidth]{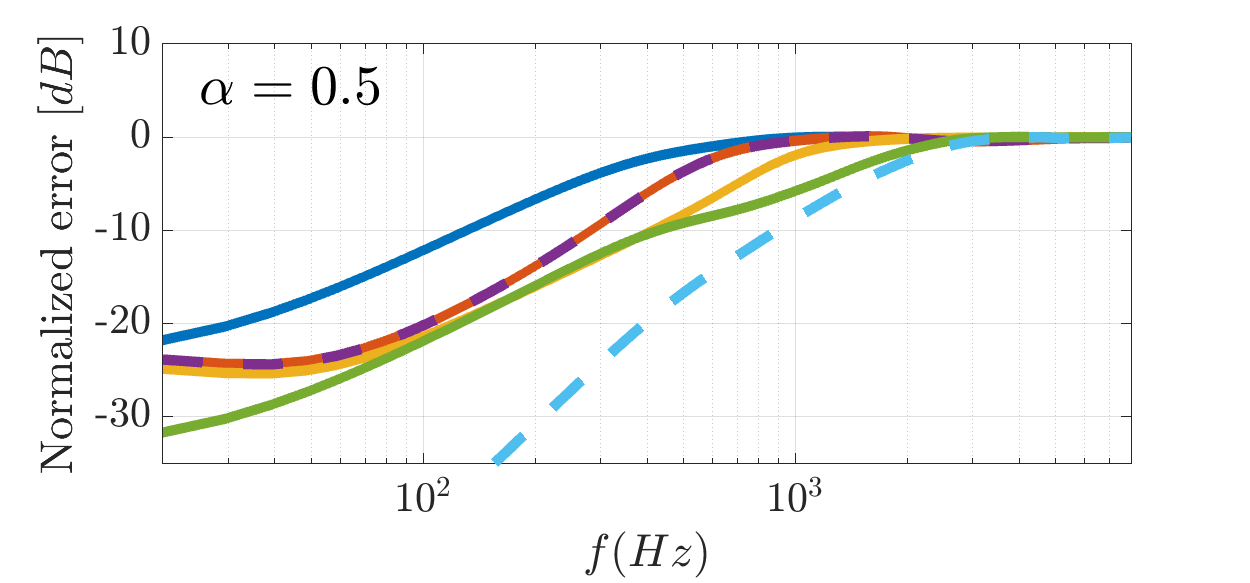}
    \caption{Normalized errors for the case of $\alpha=1$ (top), showing $\xi_{nm}^{\text{ASM}}$ for $(n,m)=(0,0)$ to $(1,1)$, $\xi_{\mathbf{nm}}^{\text{BSM}}$ and $\xi^{\text{BSM}}$. The same errors are shown for $\alpha=0$ (middle) and $\alpha=0.5$ (bottom).}
    \label{fig:SemiCircbin-joint}
\end{figure}

Fig. \ref{fig:SemiCircbin-joint} shows the performance of the joint ASM-BSM design. For $\alpha=1$ (upper), the Ambisonics representation is optimized, exhibiting minimal errors at low frequencies which rises only above $1$ kHz. However, the binaural error is considerably high for this method. In contrast, for $\alpha=0$ (middle), the binaural matching is near optimal with errors comparable to those of the standard BSM method while the Ambisonics reconstruction incurs larger errors. For the intermediate case of $\alpha=0.5$ (lower), the joint ASM-BSM design strikes a balance: Ambisonics errors are reduced relative to the the case of $\alpha=0$, but are higher than in the $\alpha=1$ case. On the other hand, the binaural errors show only a slight increase compared to the $\alpha=0$ case. Overall, these results highlight the intrinsic tradeoff between high-fidelity Ambisonics reconstruction and effective binaural signal matching.

\section{Conclusion and Future Work}
A framework for designing an Ambisonics encoder for a wearable array was presented. Matching for Ambisonics in this case typically leads to a high binaural error, significantly higher than what can usually be achieved by BSM. The joint ASM-BSM design enables the optimization of ASM filters that are also tailored for BSM. These filters achieve a much lower BSM error but at the cost of increased Ambisonics error.
The proposed formulation has a closed-form solution and provides a trade-off between ASM and BSM. Currently, only a single listener head orientation relative to the array is considered, and so future research should investigate the effects of head rotation and head tracking. Additionally, future work could explore this approach with more realistic HRTFs and conduct listening tests to assess the advantages of the joint ASM-BSM design. It would also be valuable to understand the implications of errors in the ambisonics signals on applications beyond binaural reproduction.

\bibliography{main}

\end{document}